# Clinically Translatable Direct Patlak Reconstruction from Dynamic PET with Motion Correction Using Convolutional Neural Network

Nuobei Xie[1†], Kuang Gong[2†], Ning Guo[2], Zhixing Qin[3], Jianan Cui[1], Zhifang Wu[3], Huafeng Liu[1(✉)], Quanzheng Li[2]

[1] Zhejiang University, Hangzhou, China
liuhf@zju.edu.cn
[2] Department of Radiology, Massachusetts General Hospital and Harvard Medical School, Boston, USA
[3] First Hospital of Shanxi Medical University, Taiyuan, China

**Abstract.** Patlak model is widely used in $^{18}$F-FDG dynamic positron emission tomography (PET) imaging, where the estimated parametric images reveal important biochemical and physiology information. Because of better noise modeling and more information extracted from raw sinogram, direct Patlak reconstruction gains its popularity over the indirect approach which utilizes reconstructed dynamic PET images alone. As the prerequisite of direct Patlak methods, raw data from dynamic PET are rarely stored in clinics and difficult to obtain. In addition, the direct reconstruction is time-consuming due to the bottleneck of multiple-frame reconstruction. All of these impede the clinical adoption of direct Patlak reconstruction. In this work, we proposed a data-driven framework which maps the dynamic PET images to the high-quality motion-corrected direct Patlak images through a convolutional neural network. For the patient's motion during the long period of dynamic PET scan, we combined the correction with the backward/forward projection in direct reconstruction to better fit the statistical model. Results based on fifteen clinical $^{18}$F-FDG dynamic brain PET datasets demonstrates the superiority of the proposed framework over Gaussian, nonlocal mean and BM4D denoising, regarding the image bias and contrast-to-noise ratio.

## 1 Introduction

Positron emission tomography (PET) plays an important role in neurology [1], cardiology [2] and oncology [3] studies. Compared with static PET, dynamic PET incorporates additional temporal information of tracer kinetics [4-7], which is of significance for tumor staging, tissue metabolic estimation, and treatment monitoring [8, 9]. Patlak model [10-13] is a widely used graphic model for irreversible tracers, e.g. $^{18}$F-FDG. Conventionally, the Patlak graph plot is indirectly estimated through a two-step procedure: firstly dynamic PET images are reconstructed, and then the reconstructed PET series will be fit to the Patlak model based on least-squares estimation. Although the indirect methods are overall simple to implement [14], the noise distribution is not correctly modelled as the reconstructed PET images does not follow any simple distribution. In comparison, direct Patlak reconstruction methods combine the image reconstruction and Patlak modelling in a united framework, directly from PET raw data (sinogram) [15]. Due to better noise model and more information extracted from the raw data, direct methods can reconstruct Patlak images with higher quality.

---
† indicates equal contributions.

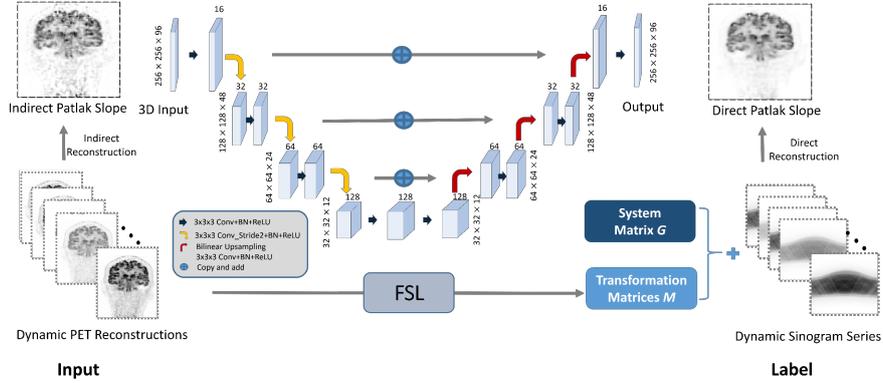

**Fig. 1.** The overall flow-chart of proposed method. The neural network is based on 3D U-Net. On the input side, we have the indirect Patlak slope computed from dynamic PET reconstructions. On the label side, the training label is the direct Patlak slope computed from raw sinogram data and motion corrected system matrices.

However, a lot of challenges still exist for the clinical adoption of direct Patlak reconstruction: 1) most researchers, especially clinicians, can only access and process the reconstructed PET images, rather than raw dynamic data. 2) Compared to indirect method, direct Patlak reconstruction is much more time-consuming as multiple frames need to be reconstructed in one framework, which is not clinically feasible. 3) Dynamic PET scan usually takes more than 60 minutes, while it is unbearable for many patients to keep still. In that sense, unavoidable motion exists in the dynamic series.

In this work, we proposed a data-driven approach which can directly compute the motion-corrected direct Patlak images from the indirect reconstructions. A modified 3D U-Net model [16, 17] was adopted as the network structure. Regarding motion correction, firstly, the transformation matrices were derived using FSL [18, 19] and incorporated with the backward/forward projectors during direct Patlak reconstruction. Fifteen clinical $^{18}$F-FDG dynamic brain PET datasets were used for evaluation of the proposed framework. Through this proposed framework, the high-quality motion corrected Patlak images can be derived in seconds solely based on clinically-accessible dynamic PET images, which is much more feasible for clinical translation.

## 2 Method

### 2.1 Overall Framework

As the framework demonstrated in Fig. 1, in the training procedure, we firstly adopted the raw sinogram to compute the dynamic reconstructions through ML-EM method. As conventional method does, the 3D indirect Patlak slope images can be computed as the U-Net input in Fig. 1.

Given the reconstructed series, the FSL toolbox [18] was employed to compute the transformation operator $M$ for the existing motion during the one hour scan. For the label side of Fig. 1, the motion corrected 3D Patlak images were directly reconstructed from the raw sinogram data, through the combination of the system matrix $G$ and transformation operator $M$. After the 3D U-Net's being training using multiple datasets, the high-quality Patlak images can be estimated from the dynamic PET frames directly in seconds without the need of raw data.

### 2.2 PET Reconstruction with Motion Correction

In general dynamic PET model, given the measured data $y = [y_1, y_2, \ldots, y_k, \ldots y_T] \in \mathbb{R}^{L \times T}$, where $y_k \in \mathbb{R}^{L \times 1}$ denotes the sum of the collected photons in PET detectors at the $k$-th time frame, the reconstruction procedure can be modeled as the affine transform

$$\bar{y}_k = G x_k + s_k + r_k \ , \tag{1}$$

where $\bar{y}_k$ denotes the expectation of $y_k$; $x_k \in \mathbb{R}^N$ represents the $k$-th image to be recovered; $G \in \mathbb{R}^{L \times N}$ is the system matrix; $s_k$ and $r_k$ are error terms caused by scatter and random events respectively. Here $L$ is the number of lines of response (LOR) and $N$ is the number of voxels in $x_k$. Conventionally, the maximum-likelihood expectation-maximization (ML-EM) update can be written as

$$x_k^{n+1} = \frac{x_k^n}{G' \mathbf{1}_L} G' \frac{y_k}{G x_k^n + s_k + r_k} \ . \tag{2}$$

Here $\mathbf{1}_L$ denotes the all 1 vector of length $L$; $G'$ denotes the transpose of $G$.

When combined with the motion correction, equation (1) can be modified as

$$\bar{y}_k = G M_k'(x_{k,mc}) + s_k + r_k \ , \tag{3}$$

where $M_k'$ denotes as inverse operator of $k$-th transformation operator $M_k$; $x_{k,mc}$ denotes the motion corrected reconstruction at $k$-th time frame. Then the motion corrected reconstruction iterates based on ML-EM is

$$x_{k,mc}^{n+1} = \frac{x_{k,mc}^n}{M_k(G' \mathbf{1}_L)} M_k(G' \frac{y_k}{G M_k'(x_{k,mc}^n) + s_k + r_k}) \ . \tag{4}$$

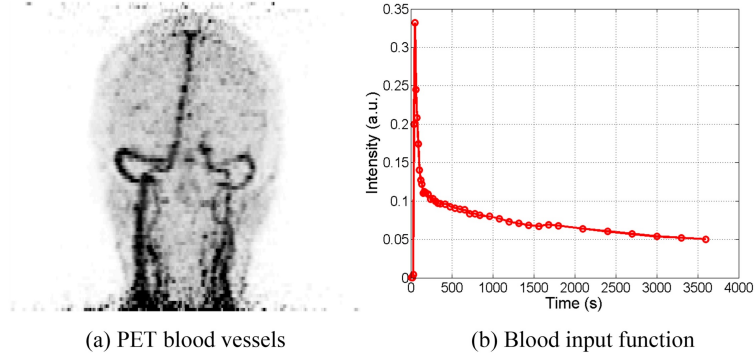

(a) PET blood vessels  (b) Blood input function

**Fig. 2.** Extract the blood input function $C_P(t)$ from each dynamic scan. (a) The maximum intensity projection (MIP) images were adopted to plot blood vessels in head. The masks for neck arterial vessels were extracted from the corresponding regions. (b) The blood input function $C_P(t)$ was exploited from the arterial activity across the scan time.

### 2.3 Direct Patlak Reconstruction

Patlak graphical model [10] is a classic linear kinetic model in dynamic PET study. In this model, the tracer concentration $C_T(t)$ at time $t$ can be written as:

$$C_T(t) = \kappa \int_0^t C_P(\tau)\, d\tau + b C_P(t), t > t^*, \qquad (5)$$

where $C_P(t)$ represents the blood input function, as demonstrated in Fig. 2; $t^*$ represents the time when the kinetic model reaches the steady state; $\kappa \in \mathbb{R}^N$ and $b \in \mathbb{R}^N$ correspondingly represent the Patlak slope and Patlak intercept. The Patlak slope $\kappa$ represents the overall influx rate of the tracer into the irreversible compartment and has found applications in many studies. From physiological perspective, the $k$-th PET frame $x_k$ in (1) can be also expressed as

$$x_k = \int_{t_s}^{t_e} C_T(\tau) e^{-\lambda \tau} d\tau, \qquad (6)$$

where $t_s$ and $t_e$ are the start and end time for the $k$-th time frame; λ denotes the decay constant. Combined with (5), equation (6) can be rewritten as

$$x_k = \int_{t_s}^{t_e}(\kappa \int_0^t C_P(\tau_1)\, d\tau_1 + bC_P(t))e^{-\lambda \tau} d\tau = B_1(k)\kappa + B_2(k)b. \qquad (7)$$

Here $B_1(k) = \int_{t_s}^{t_e} \int_0^t C_P(\tau_1)\, d\tau_1\, e^{-\lambda \tau} d\tau$ and $B_2(k) = \int_{t_s}^{t_e} C_P(t) e^{-\lambda \tau} d\tau$ serve as the basis functions. Therefore, given the dynamic PET images $x \in \mathbb{R}^{N \times T}$ in indirect reconstruction, the estimation for Patlak slope $\kappa$ and intercept $b$ can be generalized as typical linear regression task. In this work, the indirect input was computed by least-squares method.

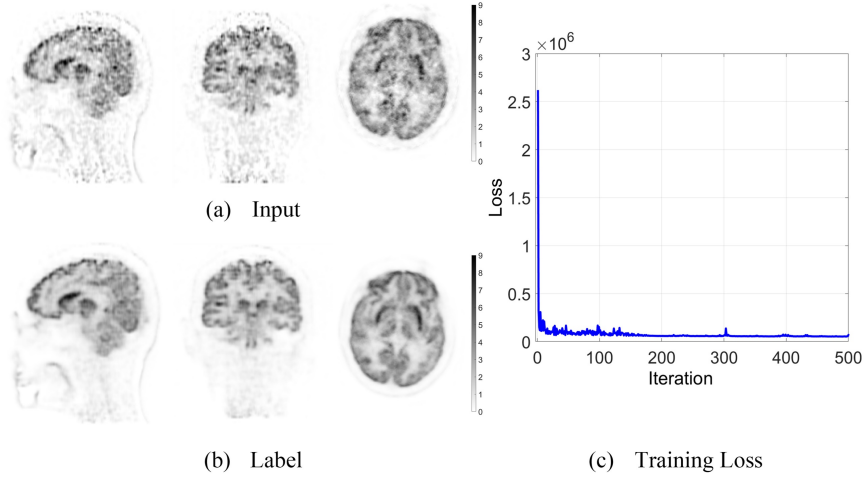

**Fig. 3.** The Patlak images from Patient 1 and corresponding training loss. (a) The input data were computed by indirect Patlak reconstruction. (b) The label data were computed by motion corrected direct Patlak reconstruction. (c) The training loss for the 3D U-Net model.

In the proposed direct Patlak reconstruction, the motion corrected Patlak parameter $\boldsymbol{\theta} = [\boldsymbol{\kappa}', \boldsymbol{b}']' \in \mathbb{R}^{2 \times N}$ can be directly reconstructed from dynamic raw sinogram series $\boldsymbol{y}$:

$$\boldsymbol{y} = (\boldsymbol{I}_T \otimes \boldsymbol{G})(\boldsymbol{M}'(\boldsymbol{x}_{mc})) + \boldsymbol{s} + \boldsymbol{r} \ , \tag{8}$$

$$\boldsymbol{x}'_{mc} = \boldsymbol{B}\boldsymbol{\theta} \ . \tag{9}$$

Here $\boldsymbol{I}_T$ denotes a $T \times T$ sized identity matrix; $\otimes$ denotes the Kronecker product; $\boldsymbol{B} \in \mathbb{R}^{T \times 2}$ denotes the collection of the basis function for all $T$ frames; $\boldsymbol{x}_{mc} \in \mathbb{R}^{N \times T}$ denotes the motion corrected images series for dynamic PET.

The nested EM [14] algorithm was used to solve the proposed direct Patlak reconstruction framework, which essentially consists of two-step EM procedures in accordance with equation (8) and (9)

$$\boldsymbol{x}_{mc}^{n+1} = \frac{\boldsymbol{x}_{mc}^n}{\boldsymbol{M}(\boldsymbol{I}_T \otimes \boldsymbol{G})' \mathbf{1}_{LT}} \boldsymbol{M}\left[(\boldsymbol{I}_T \otimes \boldsymbol{G})' \frac{\boldsymbol{y}}{(\boldsymbol{I}_T \otimes \boldsymbol{G})\boldsymbol{M}'(\boldsymbol{x}_{mc}^n) + \boldsymbol{s} + \boldsymbol{r}}\right] \ , \tag{10}$$

$$\boldsymbol{\theta}^{n+1} = \frac{\boldsymbol{\theta}^n}{\boldsymbol{B}' \mathbf{1}_{NT}} \boldsymbol{B} \frac{(\boldsymbol{x}_{mc}^{n+1})'}{\boldsymbol{B}' \boldsymbol{\theta}^n} \ , \tag{11}$$

Where $\mathbf{1}_{LT}$ and $\mathbf{1}_{NT}$ respectively denote the all one vectors in length $LT$ and $NT$. Given the significance in clinical analysis, the Patlak slope image $\boldsymbol{\kappa}$ was mainly adopted and analyzed in this study. In practice, the attenuation and normalization should also be considered [19].

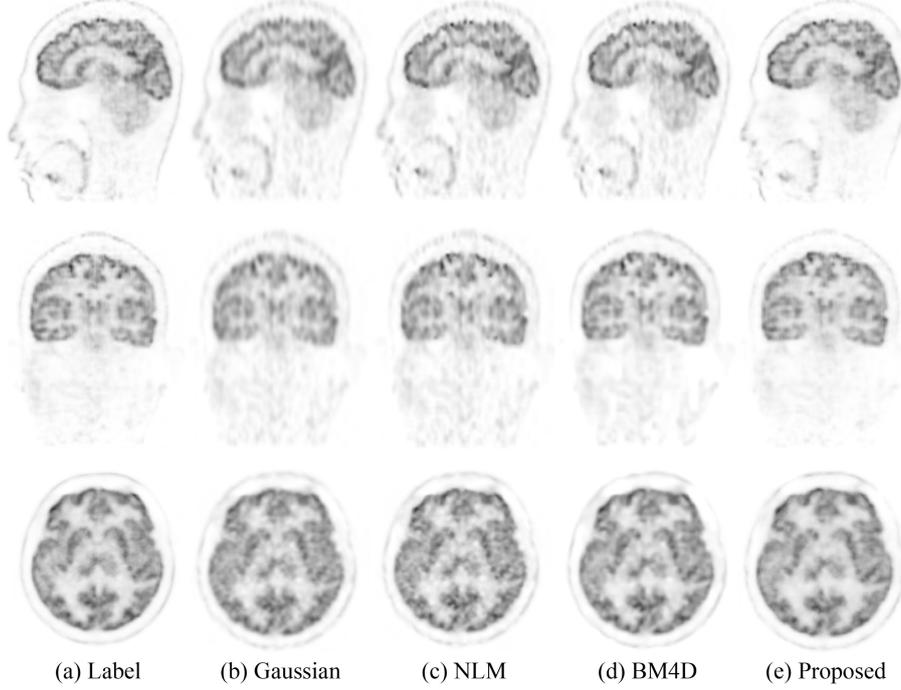

(a) Label  (b) Gaussian  (c) NLM  (d) BM4D  (e) Proposed

**Fig. 4.** Three orthogonal views of Patient 4's Patlak slope image $\kappa$. (a) Label (CNR=39.67) (b) Gaussian filtering (CNR=22.08) (c) nonlocal means (CNR=24.71) (d) BM4D (CNR=30.81) (e) Proposed (CNR=**40.32**)

## 3  Experiments

### 3.1  Data Preprocessing and Experimental Implementation

This dataset consists of 15 subjects of 60-minute $^{18}$F-FDG dynamic brain PET scan, with 42 frames for each patient: $6 \times 10s, 8 \times 15s, 6 \times 30s, 8 \times 60s, 8 \times 120s, 6 \times 300s$. All the data were acquired by the 5-ring GE Discovery MI PET/CT scanner. In reconstruction, the image size is $256 \times 256 \times 89$, with the voxel size of $1.1719 \times 1.1719 \times 2.8$ mm$^3$. For the down/up-sampling purposes, zero-padding was adopted to translate the image into $256 \times 256 \times 96$. Fig. 3(a)(b) demonstrate the input data and the label data for one of the patients.

In this study, a 5-layer 3D U-Net was adopted as the network structure which was implemented in TensorFlow 1.8. The overall structure is demonstrated in Fig. 1. In this model, the operational layers consist of : 1) $3 \times 3 \times 3$ convolutional layer; 2) batch normalization (BN) layer; 3) Relu layer; 4) $3 \times 3 \times 3$ stride-2 convolutional layer as the down sampling layer and 5) bilinear interpolation layer as the upsampling layer [17]. Besides, instead of using the concatenation operator, copy and add were employed to connect the downsampling and upsampling branches, for the purpose of reducing the parameters. The model was trained using Adam optimizer [20] and $l_2$ norm served for the cost function between the label and output 3D images. Fig. 4(c) demonstrates the training loss for the dataset. Here all the networks were trained and tested on Nvidia RTX 2080 Ti. In this study, five-fold cross-validation was conducted: 3 patients as the test set and 12 patients as the training set.

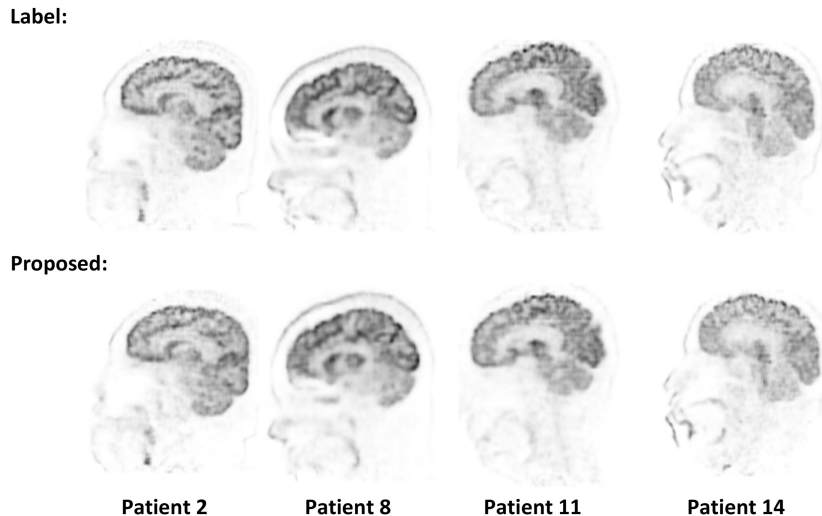

**Fig. 5.** Patlak slope images $\kappa$ from other 4 validation datasets. The structure and contrast are fully recovered in proposed method.

Given the fact that no ground truth is available in the real data study, the contrast to noise ratio (CNR) [21] was adopted as the quantitative evaluation:

$$\text{CNR} = \frac{(m_{ROI} - m_{background})}{SD_{background}} . \quad (12)$$

Here the $SD_{background}$ and $m_{background}$ respectively denote the mean value and the standard deviation of the background region (white matter in this brain PET study); $m_{ROI}$ denotes the mean value of the region of interest (ROI, gray matter). Normally, higher CNR stands for lower noise level and more distinctive contrast. In this study, the results of proposed method are compared with the label data and results from 4 state-of-the-art algorithms: Gaussian filtering, nonlocal means (NLM) [22], and BM4D methods [23]. In addition, bias between label and images from different methods was also calculated.

### 3.2 Results

Fig. 4 compares three orthogonal views of Patient 4's Patlak slope images $\kappa$ for different methods. According to the figure, the proposed method manages to reconstruct

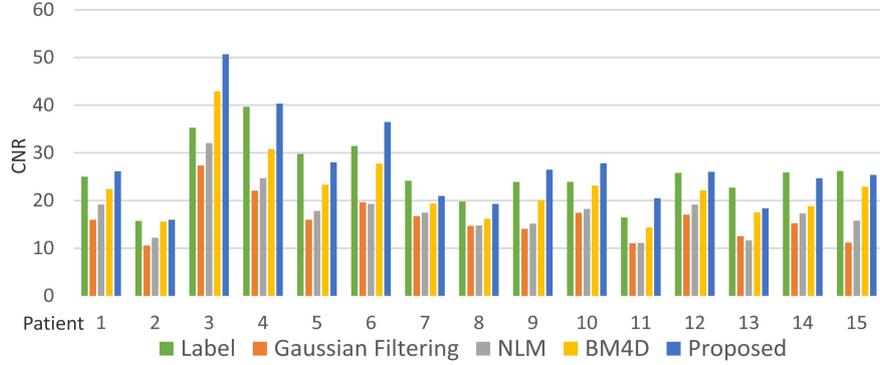

**Fig. 6.** CNR for 15 validation datasets. The proposed method demonstrates superior CNR over the counterpart of other algorithms.

the Patlak image in a comparable quality with that of the label (direct Patlak reconstructed).

In comparison with gaussian filtering and nonlocal means method, the proposed method shows superiority in preserving the contrast and also reducing the noise; when compared with BM4D results, the proposed method prevents the unusual dark/white artifacts which are derived from the noisy input. As shown in Fig. 5, for patients in other 4 exemplar datasets, the proposed method recovered comparable contrast and structures as compared with the label data.

Moreover, the quantitative results for 15 validation datasets were demonstrated in Fig. 6. For the validation datasets, the proposed method not only has better CNR compared with other compared methods, but also has competitive results with regard to the label data, achieving even higher CNR in some cases. This performance partially attributes to the denoising property inherited from the coding/decoding structure of the 3D U-Net, which is also discussed in recent deep image prior (DIP) works [24-26]. In addition, the motion-corrected direct Patlak slope images were adopted as the ground truth to calculate the bias for each method, with results shown in Fig. 7. It can be seen that the proposed method has the minimum bias compared with other methods. The computation-time for different methods is listed in Table 1. We can tell that the proposed method is much faster than the BM4D method and the original motion-corrected direct reconstruction method.

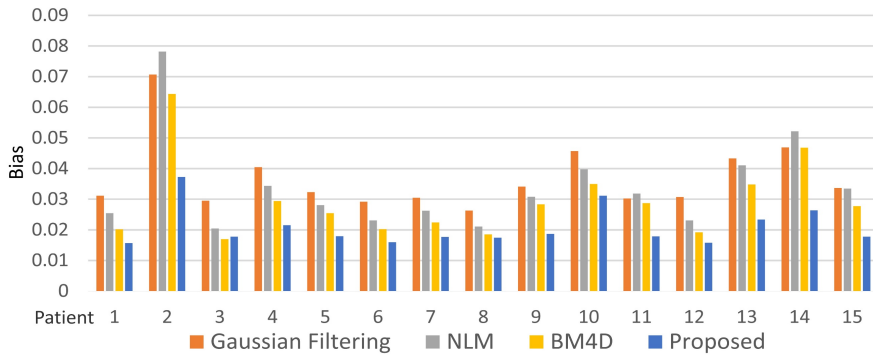

**Fig. 7.** The bias between label and results of different methods for 15 validation datasets.

Table 1. The computational time for each method

| Method | Direct Reconstruction (Label) | Gaussian Filtering | NLM | BM4D | Proposed |
|---|---|---|---|---|---|
| Computational time (s) | 7134.2 | 0.7 | 33.4 | 723.7 | 8.9 |

## 4 Conclusion

The proposed method provides a clinical translatable approach to apply motion correction and direct Patlak reconstruction for dynamic positron emission tomography (PET), based on convolutional neural network. Based on the experiments on dynamic $^{18}$F-FDG PET datasets, robust and high-quality parametric images can be estimated in seconds from dynamic PET images without raw data. Future work will focus on modifying the network structure and improving the original direct Patlak reconstruction to achieve better performance.

**Acknowledgements.** This work is supported in part by the National Natural Science Foundation of China (No: 61525106, 61427807, U1809204), by the National Key Technology Research and Development Program of China (No: 2017YFE0104000，2016YFC1300302）.